\documentclass[doublecol]{epl2}
\usepackage{bm}
\usepackage{amsmath}
\usepackage{cases}
\usepackage{appendix}
\makeatletter
\renewcommand{\maketag@@@}[1]{\hbox{\normalsize\normalfont#1}}%
\usepackage[colorlinks,
linkcolor=red,
anchorcolor=green,
citecolor=blue
]{hyperref}
\makeatother

\title{Topological $p_z$-wave nodal-line superconductivity with flat surface bands in the AH$_{\bm x}$Cr${_{\bm 3}}$As${_{\bm 3}}$ (A=Na, K, Rb, Cs) superconductors}
\shorttitle{Title} %Insert here a short version of the title if it exceeds 70 characters

\author{Juan-Juan Hao \inst{1}\footnote{These two authors contributed equally to this work.} \and Ming Zhang \inst{1}\footnote{These two authors contributed equally to this work.} \and Xian-Xin Wu \inst{2,3}\and Fan Yang \inst{1}\footnote{Email:yangfan\_blg@bit.edu.cn}}
\shortauthor{J Hao, M Zhang, X Wu and F Yang}

\institute{
	\inst{1} School of Physics, Beijing Institute of Technology, Beijing 100081, China \\
    \inst{2} CAS Key Laboratory of Theoretical Physics, Institute of Theoretical Physics, Chinese Academy of Sciences, Beijing 100190, China \\
    \inst{3} {Max-Planck-Institut f\"ur Festk\"orperforschung, Heisenbergstrasse 1, D-70569 Stuttgart, Germany}
	
}

\abstract{We study the pairing symmetry and the topological properties of the hydrogen-doped ACr$_3$As$_3$ superconductors. Based on our first-principle band structure with spin-orbit-coupling (SOC), we construct tight-binding model including the on-site SOC terms, equipped with the multi-orbital Hubbard interactions. Then using the random-phase-approximation (RPA) approach, we calculate the pairing phase diagram of this model. Our RPA results yield the triplet $p_z$-wave pairing in the $\left(\uparrow\downarrow+\downarrow\uparrow\right)$ spin channel to be the leading pairing symmetry all over the experiment relevant hydrogen-doping regimes. This pairing state belongs to the spin-U(1)-symmetry protected time-reversal-invariant topological nodal-line superconductivity. Determined by the momentum-dependent topological invariant $Z\left(k_x,k_y\right)$, the whole $(001)$ surface Brillouin zone is covered with topological flat bands with different regimes covered with different numbers of flat surface bands, which can be detected by the scanning tunneling microscope experiments.}
\begin{document}

\maketitle

	\section{Introduction}
In recent years, the quasi-one-dimensional (Q1D) Cr-based superconductors have attracted great research interests. There are two Q1D Cr-based superconductors families, i.e. the A$_2$Cr$_3$As$_3$ (abbreviated as the 233) family (A=Na, K, Rb, Cs)~\cite{Mu:034803,Bao:011013,Pang:220502(R),Zhi:147004,Adroja:134505,Kong:020507(R),Balakirev:220505,Wang:01688,Zhao:11577,Taddei:187002,Bhattacharyya:05492,Wang:1-8,Taddei:180506,Zhang:37003,Jiang:16054,Miao:205129,Zhong:227001,Yang:01336v1,Pang:84-87,Tang:020506(R),Yang:147002,Tang:16-20,Adroja:044710,Wang:020508(R),Luo:047001,Wu:057401,Wu:104511,Zhou:208,Wu:07451} and the ACr$_3$As$_3$ (abbreviated as the 133) family~\cite{Bao:180404(R),Tang:543,Mu:140504(R),Liu:27006,Cao:235107,Feng:174401,Liang:214512,Cuono:214406,Liu:094511,Galluzzi:05032}, with both sharing the same Q1D alkaline-cations-separated [(Cr$_3$As$_3$)$^{2-}$]$_{\infty}$ double-walled subnanotubes (DWSN) structure. The 233 family were synthesized earlier, and have shown unconventional pairing nature~\cite{Bao:011013,Tang:020506(R),Tang:16-20,Pang:220502(R),Zhi:147004,Adroja:134505,Yang:147002,Balakirev:220505,Cao:591,Adroja:044710,Luo:047001} with possible triplet pairing symmetry\cite{Bao:011013,Tang:020506(R),Tang:16-20,Yang:147002,Cao:591,Luo:047001,Chen:012503,Yang:01336v1}. The 133 family were synthesized slightly later, obtained by deintercalating half of the alkali metal atoms with ethanol\cite{Bao:180404(R),Tang:543,Mu:140504(R),Liu:27006}. The primitive ACr$_3$As$_3$ have a spin glass ground state~\cite{Bao:180404(R),Tang:543,Feng:174401}, which is consistent with the relaxation behavior of density functional theory (DFT) calculations\cite{Xing:174508,Feng:174401,Cao:235107,Galluzzi:05032} and the muon spin-rotation ($\mu$SR) experiments~\cite{Adroja:134505}. However, when enough amount of hydrogen atoms are intercalated in the material, superconductivity (SC) emerges~\cite{Taddei:220503(R),Wu:155108,Xiang:114802,Xiang:124802}, with highest $T_c$ around 8 K\cite{Liang:214512,Liu:094511}. Therefore, the correct chemical formula for``ACr$_3$As$_3$'' should be AH$_x$Cr$_3$As$_3$. It is the difference in H content that leads to the difference in ground state properties between superconducting and non-superconducting samples~\cite{Taddei:220503(R),Wu:155108,Xiang:114802,Xiang:124802}. The proximity of the magnetic ordered state to the SC in the phase diagram implies spin fluctuation-driven pairing mechanism, similar to the cuprates and the iron-based superconductors.

\begin{figure*}
		\centering
		\includegraphics[width=1\textwidth]{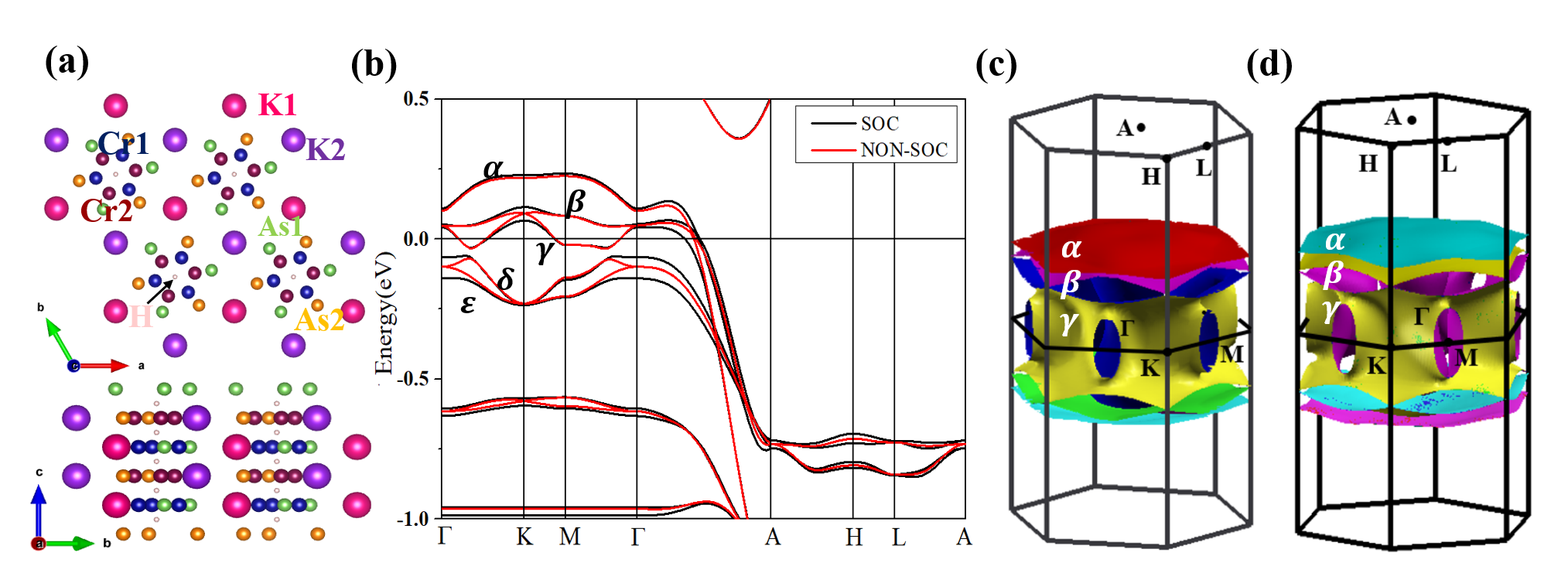}
		%	\onefigure[width=1\textwidth]{band_structrue.pdf}
		\caption{(Color online). Crystal and band structures of KHCr$_3$As$_3$. (a) Top and side views for the crystal structure. (b) Band structures with (black) and without (red) SOC along the high-symmetry lines. The five bands near the Fermi level are marked as $\alpha$ - $\varepsilon$ respectively. (c) and (d) are FSs without and with SOC, with the high symmetric points marked in the 3D BZ.}
		\label{fig.1}
	\end{figure*}

The density-functional theory (DFT) based calculations suggest that the low-energy degrees of freedom for the AH$_x$Cr$_3$As$_3$ is the Cr-3d orbitals\cite{Taddei:220503(R),Wu:155108,Xiang:114802,Xiang:124802,Zhang:14367}, similar with those of the 233 family\cite{Bao:011013,Tang:020506(R),Tang:16-20,Pang:220502(R),Kong:020507(R),Wang:020508(R),Taddei:180506,Luo:047001,Jiang:16054,Wu:057401,Wu:104511,Zhang:37003,Zhou:208,Miao:205129,Zhong:227001,Wu:07451}, implying the strong electron-electron interactions can account for the origin of the SC in the system. Furthermore, the presence of H induces effective electron doping, which confirms that H doping has the effect of raising Fermi level~\cite{Taddei:220503(R),Zhang:14367,Wu:155108}. A remarkable property of the band structure of the AH$_x$Cr$_3$As$_3$ is the presence of the Lifshitz transition under moderate hole doping, during which its 3D FSs are changed to two disconnected Q1D FS sheets~\cite{Wu:155108,Zhang:14367}. It's clarified in our previous work\cite{Zhang:14367} based on the random-phase approximation (RPA)~\cite{Yada:2161,Kuroki:087004,Kubo:224509,Graser:025016,Maier:100515(R),Liu:217001,Liu:066804,Takimoto:104504,Zhang:094511} that the presence of the type-II VHS\cite{Yao:035132,Ma:245114,Meng:184509,Chen:174503,Zhang:104504} during the Lifshitz transition doping regimes has driven the triplet $p_z$-wave pairing symmetry. It's interesting to investigate the topological properties of this triplet pairing state. However, as the spin-orbit coupling (SOC) has not been considered in that work, the three spin components of the triplet pairing are exactly degenerate, which leads to uncertainty of the superconducting ground state.

In this paper, we report our RPA study on the pairing symmetry and the topological properties of the KH$_x$Cr$_3$As$_3$, based on our first-principle DFT band structure considering SOC. Our main results are as follow. Due to the inversion symmetry of the material, the double-folded spin degeneracy survives the SOC. The main effect of the SOC on the band structure lies in the elimination of the double degeneracy along the $\Gamma$-$A$ and the $K$-$H$ lines. Starting from our previous six-orbital tight-binding (TB) model\cite{Zhang:14367} without SOC, we add the dominating on-site SOC conserving the spin-U(1) symmetry (SUS), leading to a band structure well consistent with the DFT one. Then we add the multi-orbital Hubbard interactions into this model and perform the RPA calculations. The leading pairing symmetry obtained from our RPA calculations is triplet $p_z$-wave pairing in the $S_z=0$ channel, i.e. the $\uparrow\downarrow+\downarrow\uparrow$. This pairing state belongs to the SUS-protected time-reversal-invariant (TRI) topological nodal-line SC defined in the Ref\cite{Liu:033050}, which includes the 233 family as a good example. Determined by the momentum-dependent topological winding number, the whole surface Brillouin zone (BZ) is covered with topological flat bands, with different regimes covered with different numbers of flat bands, as shown by our surface spectrum in the $(001)$ surface BZ. Our results appeal more experimental access into this intriguing family.

\section{The DFT Band Structure}

The crystal structure of KHCr$_3$As$_3$ is shown in Fig.~\ref{fig.1}(a). This compound consists of an infinite linear chain of DWSN [(Cr$_3$As$_3$)$^{2-}$]$_{\infty}$ extended along the c-axis, forming a Q1D structure, which is interconnected by K$^+$ cations. The hydrogen atoms locate at the center of the Cr$_6$ octahedrons in such a DWSN structure\cite{Taddei:220503(R),Wu:155108,Xiang:114802,Xiang:124802,Zhang:14367}. Compared with the crystal structure of K$_2$Cr$_3$As$_3$~\cite{Bao:011013,Tang:020506(R),Tang:16-20,Pang:220502(R),Kong:020507(R),Wang:020508(R),Taddei:180506,Luo:047001,Jiang:16054,Wu:057401,Wu:104511,Zhang:37003,Zhou:208,Miao:205129,Zhong:227001,Wu:07451} and KCr$_3$As$_3$~\cite{Zhang:094511,Bao:180404(R),Tang:543,Mu:140504(R),Feng:174401,Taddei:220503(R),Liu:27006,Cao:235107,Liang:214512,Cuono:214406,Liu:094511,Galluzzi:05032}, KHCr$_3$As$_3$ is more similar to KCr$_3$As$_3$, which has a centrosymmetric structure with space group $P6_{3}/m$ (No.176)) and point group $C_{6h}$.
	
The band structure of KHCr$_3$As$_3$ is calculated using DFT theory, adopting the ultra-soft pseudopotential (USPP) of Perdew-Burke-Ernzerhof (PBE)~\cite{Perdew:3865} in QUANTUM ESPRESSO (QE)~\cite{Giannozzi:465901}. We use $5\times5\times11$ sparse k-points in the entire Brillouin Zone (BZ) for self-consistent calculation. We set the truncated kinetic energy of the wave function to 60Ry and the charge density to 400Ry. The experimental lattice parameters of the crystal structure of KHCr$_3$As$_3$ were adopted: $a = 9.09481$ \AA~ and $c = 4.17703$ \AA~\cite{Taddei:220503(R)}, with only the atomic positions within the unit cell to be determined by optimization.
	
The obtained band structure of KHCr$_3$As$_3$ without considering SOC is shown in Fig.~\ref{fig.1}(b) by the red curves. It can be seen that there are five bands near the Fermi level, labeled as $\alpha, \beta, \gamma, \delta$ and $\epsilon$, respectively. The three bands across the Fermi level are the $\alpha$-, $\beta$- and $\gamma$- bands. These bands form two Q1D Fermi surfaces ($\alpha$ and $\beta$) and a 3D Fermi surface ($\gamma$), as shown in Fig.~\ref{fig.1}(c). Note that the $\beta$ and $\gamma$- Fermi surfaces (FSs) intersect at the $k_{x}=k_{y}=0, k_z=k_F$ point on the $\Gamma$-$A$ line. Compared with the band structure of KCr$_3$As$_3$~\cite{Zhang:094511,Bao:180404(R),Tang:543,Mu:140504(R),Feng:174401,Taddei:220503(R),Liu:27006,Cao:235107,Liang:214512,Cuono:214406,Liu:094511,Galluzzi:05032}, it can be found that the Fermi level rises when H is inserted, which reflects that H is used as an electron donor in KHCr$_3$As$_3$, consistent with Ref\cite{Taddei:220503(R)}.

After the band structure calculation, a six-band low-energy tight-binding model has been constructed in our previous work\cite{Zhang:14367}. Briefly, the relevant orbital degrees of freedom include the Cr-3d$_{z^2}$, 3d$_{x^{2}-y^{2}}$ and 3d$_{xy}$ orbitals from the two inequivalent Cr atoms within each unit cell, and the effective hopping parameters are obtained through the calculations of maximal localization for these orbitals using Wannier90~\cite{Mostofi:2309}. The obtained TB Hamiltonian with $C_{6h}$ point-group symmetry can be expressed in the momentum space as,
	\begin{align}
	H_{{\rm TB}}
	=\sum_{\mathbf{k}\mu\nu\sigma}h_{\mu\nu}(\mathbf{k})
	c^{\dagger}_{\mathbf{k}\mu\sigma}c_{\mathbf{k}\nu\sigma},
	\end{align}
	Here $\mu,\nu=1,\cdots, 6$ indicate the orbital-sublattice indices, containing the $d_{z^2}$, $d_{x^2-y^2}$ and $d_{xy}$ orbitals of A or B sublattices. The elements of the $h(\bm{k})$ matrix is obtained through Fourier transformation using the real-spacing hopping-integral data provided in our previous work\cite{Zhang:14367}. The band structure thus obtained is well consistent with that of the DFT at low energy near the Fermi level, see Ref\cite{Zhang:14367}.

\section{The SOC Hamiltonian}
\begin{figure}
		%\onefigure[width=0.45\textwidth]{fermi_surface.pdf}
		\includegraphics[width=0.48\textwidth]{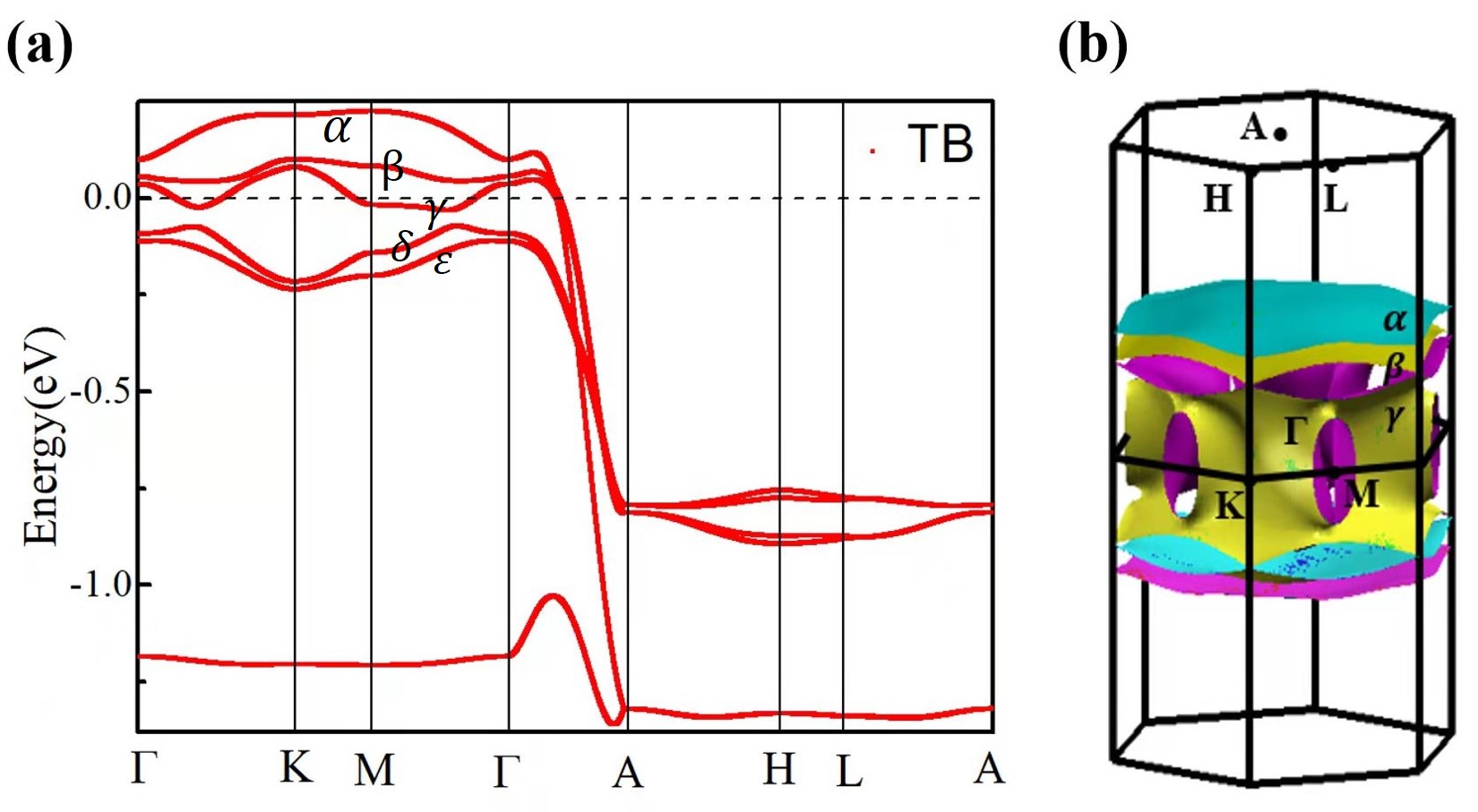}	
		\caption{(Color online). Band structure (a) and FS (b) of our band model (\ref{htb}) for KHCr$_3$As$_3$ with on-site SOC.}
		\label{band_tb_soc}
	\end{figure}

%% here a revision

Using DFT calculations, we obtain the band structure with SOC. The band structures with and without SOC are compared in Fig. ~\ref{fig.1}(b). From the comparison, we find the two band structures are overall similar except that the degeneracies along the high-symmetric lines are lift up by the SOC. Without SOC, the $\beta$- and $\gamma$- bands are degenerate at the $\Gamma$- and $K$- points, or in general along the $\Gamma$-$A$ and $K$-$H$ lines. Similarly, the $\delta$- and $\varepsilon$- bands are also degenerate along these high-symmetric lines. However, with SOC, the energies of the $\beta$- and $\gamma$- bands at the $\Gamma$- and $K$- points are split by about 0.02eV and 0.05eV, and the energies of the $\delta$- and $\varepsilon$- bands at the two momenta are split by about 0.08eV and 0.006eV, respectively. Note that the spin degeneracy is maintained with SOC, due to the combination of the inversion and time-reversal (TR) symmetries. The FS with SOC is shown in Fig. \ref{fig.1}(d). Now the $\beta$- and $\gamma$- FSs no longer intersect at the $(0,0,k_F)$ point on the $\Gamma$-$A$ line.

In the following studies involving electron-electron interactions, we need a simple TB formula for the SOC part of the Hamiltonian. For simplicity, we only consider the dominating part of the SOC, i.e. the on-site SOC term. From combination of the rotation symmetry, the mirror-reflection symmetry about the $xy$-plane, and the TR symmetry of the system, the only possible on-site SOC takes the following formula\cite{Liu:033050, Wu:104511},
\begin{eqnarray}\label{hsoc}
H_{\text{SOC}}&=&i\lambda_{\text{SOC}}\sum_{\mathbf i \sigma}\sigma [c^{\dagger}_{\mathbf i 2\sigma}c_{\mathbf i 3\sigma}-c^{\dagger}_{\mathbf i 3\sigma}c_{\mathbf i 2\sigma}]+i\lambda_{\text{SOC}}\nonumber\\&&\times\sum_{\mathbf i \sigma}\sigma[c^{\dagger}_{\mathbf i 5\sigma}c_{\mathbf i 6\sigma}-c^{\dagger}_{\mathbf i 6\sigma}c_{\mathbf i 5\sigma}].
\end{eqnarray}
This SOC Hamiltonian also satisfies the inversion symmetry $\hat{R}$ which dictates $c^{\dagger}_{\mathbf i \mu\sigma} \to c^{\dagger}_{\left(\hat{R}\mathbf i\right) \mu\sigma}$.

Our full model band Hamiltonian with SOC is given by
\begin{equation}
\label{htb}
H_{\text{band}}=H_{\text{TB}}+H_{\text{SOC}}.
\end{equation}
Setting $\lambda_{\text{SOC}}=10$ meV, the obtained band structure and FS for this model are shown in Fig.~\ref{band_tb_soc} (a) and (b) respectively. Comparing Fig. ~\ref{band_tb_soc} (a) with Fig. ~\ref{fig.1} (b), it's found that this band structure is qualitatively consistent with the DFT one with SOC. Firstly, the spin degeneracy is kept in this band structure, i.e. $\varepsilon_{\mathbf{k}\uparrow}^{\alpha}=\varepsilon_{\mathbf{k}\downarrow}^{\alpha}\equiv\varepsilon_{\mathbf{k}}^{\alpha}$, due to the combined TRS and inversion symmetry, where $\varepsilon_{\mathbf{k}\sigma}^{\alpha}$ denotes the band energy for the spin $\sigma$ and momentum $\mathbf{k}$ in the $\alpha$-th band. Secondly, the degeneracies between the $\beta$- and $\gamma$- bands and those between the $\delta$- and $\varepsilon$- bands at the $\Gamma$- and $K$- points and along the $\Gamma$-$A$ lines (as well as along the unshown $K$-$H$ lines) are lift up. Furthermore, the FS for the up or down spin shown in Fig.~\ref{band_tb_soc} (b) for this model is also qualitatively consistent with the DFT FS with SOC shown in Fig.~\ref{fig.1}(d).

\section{The RPA pairing phase diagram}
According to our previous RPA result for this material without considering SOC\cite{Zhang:14367}, the triplet $p_z$-wave pairing is the leading pairing symmetry whose three spin components are exactly degenerate. Here we consider the SOC to lift up such degeneracy. We adopt the following extended Hubbard model Hamiltonian in our study,
\begin{align}\label{model}
H=&H_{\text{band}}+H_{\text{int}}\nonumber\\
H_{\text{int}}=&U\sum_{i\mu}n_{i\mu\uparrow}n_{i\mu\downarrow}+
V\sum_{i,\mu<\nu}n_{i\mu}n_{i\nu}+J_{H}\sum_{i,\mu<\nu}                   \nonumber\\
&\Big[\sum_{\sigma\sigma^{\prime}}c^{+}_{i\mu\sigma}c^{+}_{i\nu\sigma^{\prime}}
c_{i\mu\sigma^{\prime}}c_{i\nu\sigma}+(c^{+}_{i\mu\uparrow}c^{+}_{i\mu\downarrow}
c_{i\nu\downarrow}c_{i\nu\uparrow}+h.c.)\Big]
\end{align}
Here, the interaction parameters $U$, $V$, and $J_H$ denote the intra-orbital, inter-orbital Hubbard repulsion, and the Hund's rule coupling (as well as the pair hopping) respectively, which satisfy the relation $U=V+2J_H$. Note that this Hamiltonian conserves the SUS, i.e. the total $S_z$ is a good quantum number.

\begin{figure}
	\includegraphics[width=0.45\textwidth]{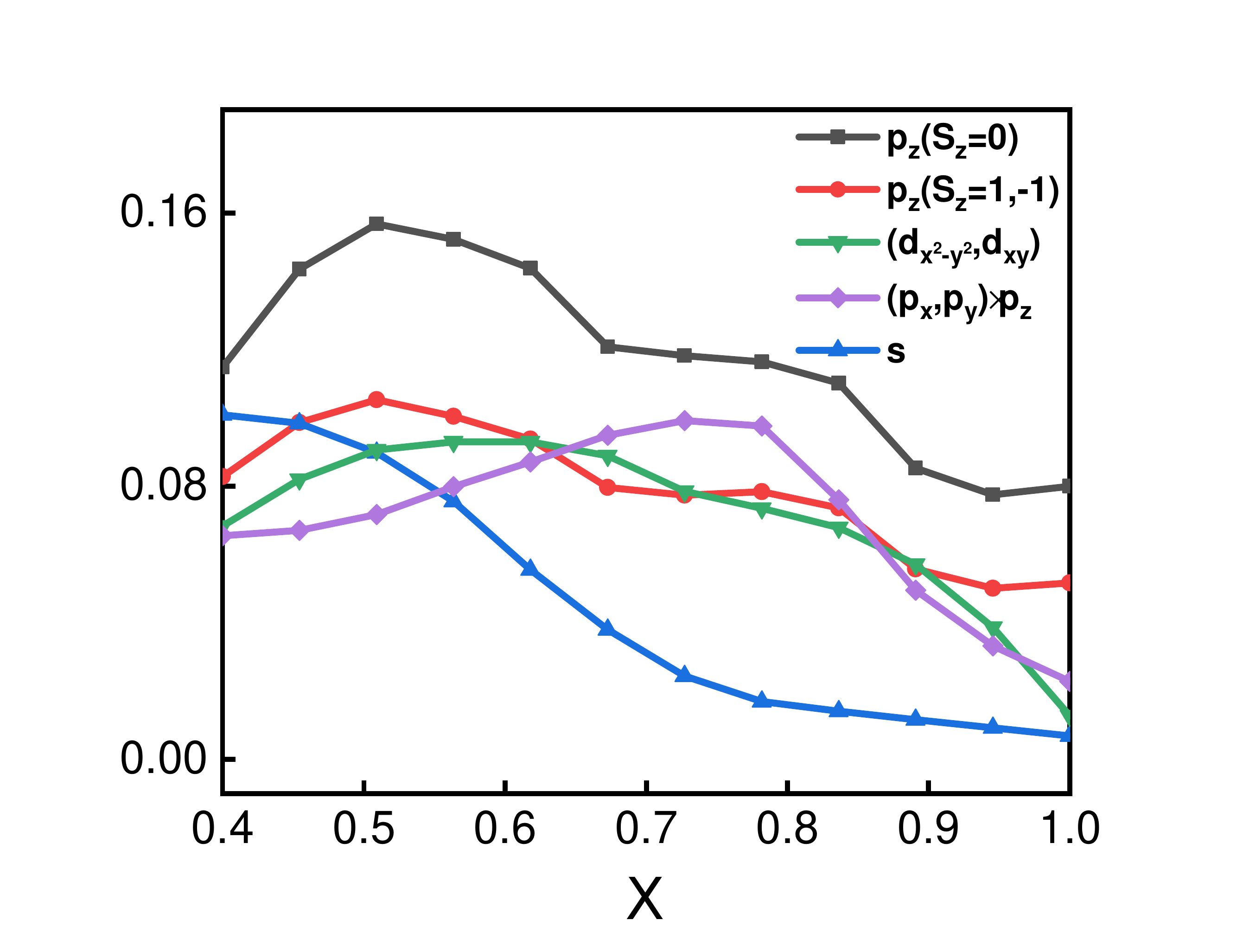}
	\caption{(Color online). Doping $x$ dependences of the largest pairing eigenvalues $\lambda$ for the $p_z$-, $(d_{x^2-y^2},d_{xy})$-, $(p_x, p_y)\times p_z$- and $s$-wave pairings for fixed $U=0.25$ eV,$J_H=0.2U$ and $\lambda_{\text{SOC}}=10$ meV. Note that for the triplet $p_z$-wave pairing, the $S_z=\pm 1$ pairing channels are degenerate, which nevertheless are non-degenerate with the $S_z=0$ channel.} \label{pzsoc}
\end{figure}

The pairing symmetry of this model is obtained through the standard multi-orbital RPA approach considering SOC\cite{Wu:104511}. The procedure of this approach is briefly introduced as follow. Firstly, we need to calculate the susceptibility tensor as function of the momentum. As the Hamiltonian (\ref{model}) conserves the SUS, the spin indices of the nonzero susceptibility tensor elements can only be $\uparrow\uparrow\uparrow\uparrow$, $\uparrow\uparrow\downarrow\downarrow$, $\downarrow\downarrow\uparrow\uparrow$, $\downarrow\downarrow\downarrow\downarrow$, $\uparrow\downarrow\downarrow\uparrow$, $\downarrow\uparrow\uparrow\downarrow$. The renormalized susceptibility tensor can be obtained from the bare one via the Dyson equation in the RPA level. Secondly, we need to calculate the effective pairing interaction mediated by exchanging the particle-hole excitations whose propagators are represented by these susceptibility tensor elements. Note that with SOC here, the effective pairing interactions in the three spin channels of the triplet pairing are no longer degenerate. Finally, we need to solve the linearized mean-field (MF) gap equation near the SC $T_c$ involving the obtained effective pairing interaction function to obtain all possible pairing eigenvalues $\lambda$ and corresponding pairing eigenvectors. The pairing eigenvectors corresponding to the largest pairing eigenvalue $\lambda$ determines the pairing symmetry, and the $T_c$ is related to $\lambda$ via $T_c\sim W_{\text{band}}e^{-1/\lambda}$, with $W_{\text{band}}$ denoting the low-energy band width.

The doping $x$ dependences of the pairing eigenvalues $\lambda$ for the five leading pairing symmetries are shown in Fig.\ref{pzsoc} in the doping regime $x\in (0.4,1)$ wherein definite experimental evidences for SC have been detected. Note that for the triplet pairings, due to the SUS, the $S_z$ of the Cooper pair is a good quantum number, which suggests that we can get the $S_z=1$ (i.e. the $\uparrow\uparrow$), $S_z=-1$ (i.e. the $\downarrow\downarrow$) or $S_z=0$ (i.e. the $\uparrow\downarrow+\downarrow\uparrow$) pairing channels in the presence of the SOC. While the $S_z=1$ and $S_z=-1$ pairing channels are still degenerate due to the spin degeneracy, they are not degenerate with the $S_z=0$ pairing channel. As a result, Fig.\ref{pzsoc} shows that the triplet $p_z$-wave pairing in the $S_z=0$ (i.e. the $\uparrow\downarrow+\downarrow\uparrow$) channel is the leading pairing symmetry. This pairing phase diagram can be compared with that without SOC provided in Ref\cite{Zhang:14367}.

\section{SUS-Protected Topological Nodal-Line SC}
The BCS-mean-field (MF) Hamiltonian describing the triplet $p_z$-wave pairing in the $S_z=0$ channel reads,
\begin{equation}\label{BCS}
\begin{aligned}
H_{\text{BCS}}&=H_{\text{TB}}+H_{\text{SOC}}+H_{\Delta},\\
H_{\Delta}&=\sum_{\mathbf {k}\alpha}\left[\left(c^{\dagger}_{\mathbf {k}\alpha\uparrow}c^{\dagger}_{\mathbf {-k}\alpha\downarrow}+
c^{\dagger}_{\mathbf {k}\alpha\downarrow}c^{\dagger}_{\mathbf {-k}\alpha\uparrow}\right)\Delta_{\mathbf {k}}^{\alpha}+h.c.\right].
\end{aligned}
\end{equation}
Here the real gap function $\Delta_{\mathbf {k}}^{\alpha}$ obtained from our RPA calculations satisfies the $p_z$-wave pairing symmetry. This symmetry requires the gap function to be zero on the $k_z=0$ plane. Therefore, the FS cut on this plane gives the gap nodal lines. It's interesting to investigate the topological properties of such nodal-line SC.

It's easily checked that the MF Hamiltonian (\ref{BCS}) satisfies the combined TRS and SUS. The topological classification of nodal-line superconductors hosting the two symmetries can be attributed to the following momentum-dependent topological invariant $Z(k_x,k_y)$ defined as\cite{Liu:033050}
\begin{equation}\label{z}
Z(k_x,k_y)=\frac{1}{2\pi}\sum_{\alpha}\int^{\pi}_{-\pi}d\theta_{\mathbf {k}\alpha}\equiv\sum_{\alpha}I_{\alpha}(k_x,k_y).
\end{equation}
Here $\theta_{\mathbf {k}\alpha}$ denotes the complex phase angle of $\varepsilon_{\mathbf {k}}^{\alpha}+i\Delta_{\mathbf {k}}^{\alpha}$, and the integer $I_{\alpha}(k_x,k_y)\equiv \frac{1}{2\pi}\int^{\pi}_{-\pi}d\theta_{\mathbf {k}\alpha}$ represents the winding number of $\theta_{\mathbf {k}\alpha}$ along a closed path in the BZ which passes through $(k_x,k_y,0)$ and is perpendicular to the $k_z=0$ plane. As topological invariant, $2Z(k_x,k_y)$ counts the number of gapless surface modes at the momentum $(k_x,k_y)$ in the $(001)$-surface BZ.

\begin{figure}
	\includegraphics[width=1.0\columnwidth]{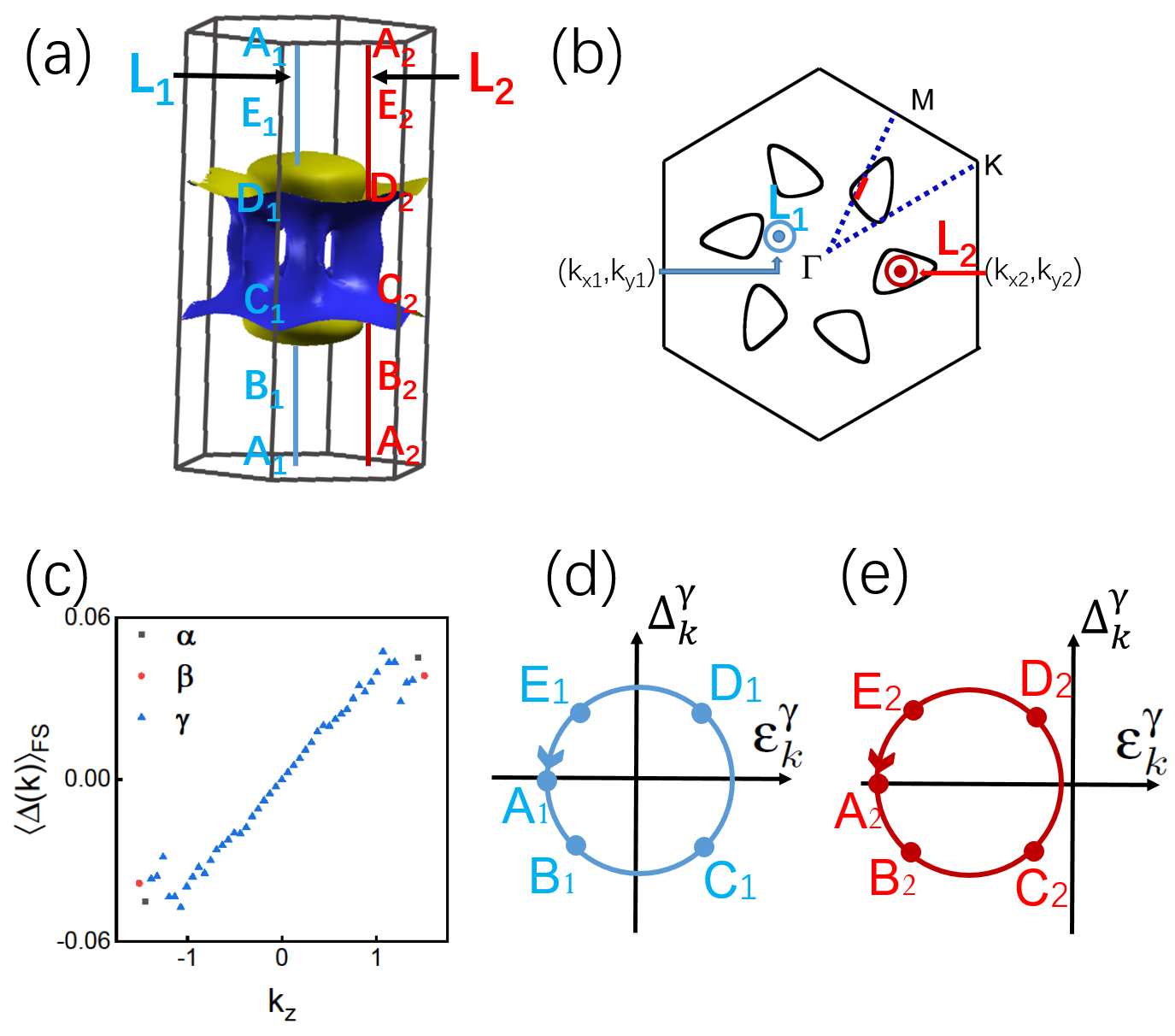}
	\caption
	{(Color online) (a) The $\gamma$-FS for one spin species considering SOC. The paths $L_{i,i=1,2}$ are perpendicular to the $(k_x,k_y)$-plane. The momenta $A_i\sim E_i$ locate on the path $L_i$. (b) The $k_z=0$ cut of the $\gamma$-FS. The path $L_1$ ($L_2$) intersects with the $k_z=0$ plane at the momentum $(k_{x1},k_{y1})$ ($(k_{x2},k_{y2})$), indicated by the center of the small blue (red) circle. The high-symmetry lines are marked. (c) The $k_z$-dependence of the relative gap function averaged on the FSs.  Schematic diagrams of how the phase angle of $\varepsilon_{\mathbf {k}\gamma}+i\Delta_{\mathbf{k}}^{\gamma}$  evolve with $k_z$ along the paths $L_1$ for (d) and $L_2$ for (e).  The doping level for (a)-(e) is $x=0.85$.}
	\label{fs}
\end{figure}

As an example, we illustrate how to calculate $I_{\gamma}(k_x,k_y)$ for two typical $(k_x,k_y)$ points at the doping level $x=0.85$. The $\gamma$- FS for $x=0.85$ is shown in Fig.~\ref{fs}(a).  This FS consists of two Q1D FS sheets nearly parallel to the $k_z=0$ plane, connected by six hollow tubes (only three are visible in the figure) nearly perpendicular to the $k_z=0$ plane. The six hollow tubes cut the $k_z=0$ plane to form six pockets, as shown in Fig.~\ref{fs}(b). The $(k_{x1},k_{y1})$ locates near the $\Gamma$- point, passed through by the closed vertical path $L_1$, and the $(k_{x2},k_{y2})$ locates within one of the six pockets, passed through by the path $L_2$. The $k_z$-dependence of $\Delta^{\alpha}_{\mathbf {k}}$ averaged on the FSs is shown in Fig.~\ref{fs}(c), where the sign of $\Delta^{\alpha}_{\mathbf {k}}$ follows that of $k_z$. This character is also satisfied for the $\Delta^{\alpha}_{\mathbf {k}}$ with fixed $(k_x,k_y)$, as the Q1D $p_z$-wave pairing gap function value doesn't obviously depend on $(k_x,k_y)$.

Let's first evaluate $I_{\gamma}$ for $(k_{x1},k_{y1})$. Fig.~\ref{fs}(a) shows the locations of five typical momenta $A_1\sim E_1$ along $L_1$, which intersects with the FS twice. The energies $\varepsilon_{\mathbf {k}}^{\gamma}$ for $C_1$ and $D_1$ are positive, and those for the remaining three are negative. The gap functions $\Delta_{\mathbf {k}}^{\gamma}$ for $D_1$ and $E_1$ are positive, and those for $B_1$ and $C_1$ are negative. Due to the $p_z$-wave symmetry, the $\Delta_{\mathbf {k}}^{\gamma}$ for $A_1$ is zero. Fig.~\ref{fs}(d) shows the locations of the $\varepsilon_{\mathbf {k}}^{\gamma}+i\Delta_{\mathbf {k}}^{\gamma}$ for the five momenta $A_1\sim E_1$ in the complex plane expanded by the $\varepsilon_{\mathbf {k}}^{\gamma}$- axis and the $\Delta_{\mathbf {k}}^{\gamma}$- axis, which form a closed trajectory circling around the original point, suggesting that the phase winding number of $\varepsilon_{\mathbf {k}}^{\gamma}+i\Delta_{\mathbf {k}}^{\gamma}$ along $L_1$ is 1. Therefore, we have $I_{\gamma}(k_{x1},k_{y1})=1$. Then let's investigate $I_{\gamma}(k_{x2},k_{y2})$. As $L_2$ locates within one of the six hollow tubes, it does not intersect with the $\gamma$-FS. Consequently, the $\varepsilon_{\mathbf {k}}^{\gamma}$ for the five momenta $A_2\sim E_2$ are all negative. On the other hand, the signs for the $\Delta_{\mathbf {k}}^{\gamma}$ for the five momenta are the same as those for $A_1\sim E_1$. As a result, the trajectory formed by the locations of the $\varepsilon_{\mathbf {k}}^{\gamma}+i\Delta_{\mathbf {k}}^{\gamma}$ for $A_2\sim E_2$ in the complex plane doesn't circle around the original point, as shown in Fig.~\ref{fs}(e), which leads to zero winding number of the phase angle of $\varepsilon_{\mathbf {k}}^{\gamma}+i\Delta_{\mathbf {k}}^{\gamma}$ along $L_2$. Thus, we have $I_{\gamma}(k_{x2},k_{y2})=0$.

\begin{figure}
	\includegraphics[width=1.0\columnwidth]{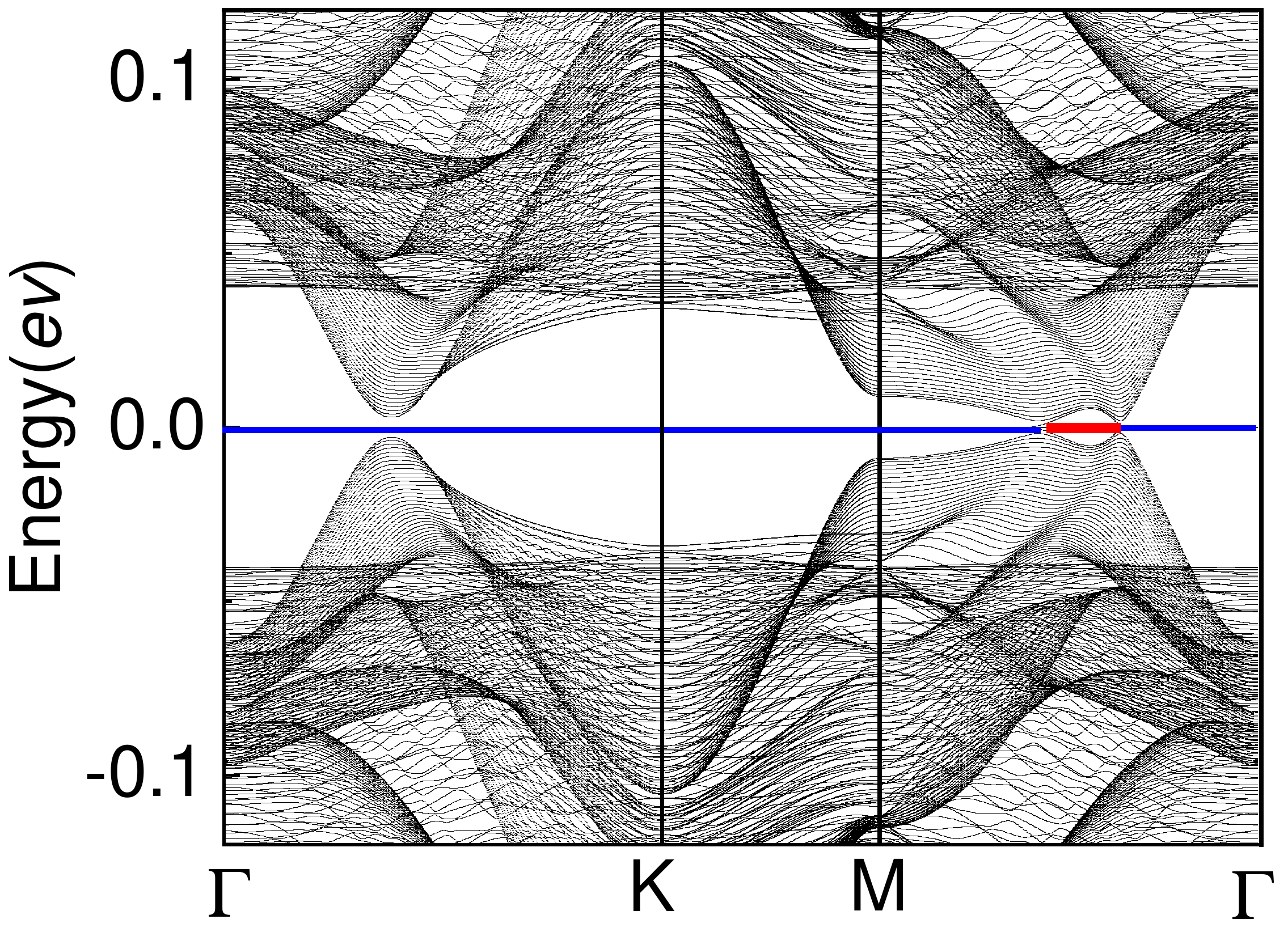}
	\caption
	{(Color online) (a) The energy spectrum for KH$_{0.85}$Cr$_3$As$_3$ along the high-symmetry lines in the $(k_{x},k_{y})$ plane, with open boundary condition along the $z$-axis and periodic ones along the $x$- and $y$- axes. The segment marked red (blue) is covered by 4 (6) flat bands. To enhance the visibility, we have enlarged the gap amplitude by an order of magnitude.}
	\label{edge}
\end{figure}

Besides the $\gamma$- FS, the compound with $x=0.85$ hosts two other Q1D FSs, i.e. the $\alpha$- and $\beta$- FSs. These two FSs take simple shapes, with each hosting two Q1D FS sheets almost parallel to the $k_z=0$ plane, similar as those for $x=1$ shown in Fig. \ref{fig.1} (d) and Fig. \ref{band_tb_soc} (b). For arbitrary $(k_x,k_y)$, the vertical path $L$ passing through it always intersects with the $\alpha$- and the $\beta$- FS twice. On the other hand, the gap functions for these two bands are similar with that for the $\gamma$ band in that the sign of $\Delta_{\mathbf {k}}^{\alpha/\beta}$ follows that of $k_z$. Following the above illustrated procedure, we can easily obtain $I_{\alpha/\beta}(k_x,k_y)=1$ for an arbitrary $(k_x,k_y)$. The total topological invariant $Z(k_x,k_y)$ is the sum of $I_{\alpha}(k_x,k_y)$, $I_{\beta}(k_x,k_y)$ and $I_{\gamma}(k_x,k_y)$, which takes different nonzero values in different regimes of the $(k_x,k_y)$ plane. For example, for the two typical momenta shown in Fig.~\ref{fs} (b), we have $Z(k_{x1},k_{y1})=3$ and $Z(k_{x2},k_{y2})=2$.
\begin{figure}
	\includegraphics[width=1.0\columnwidth]{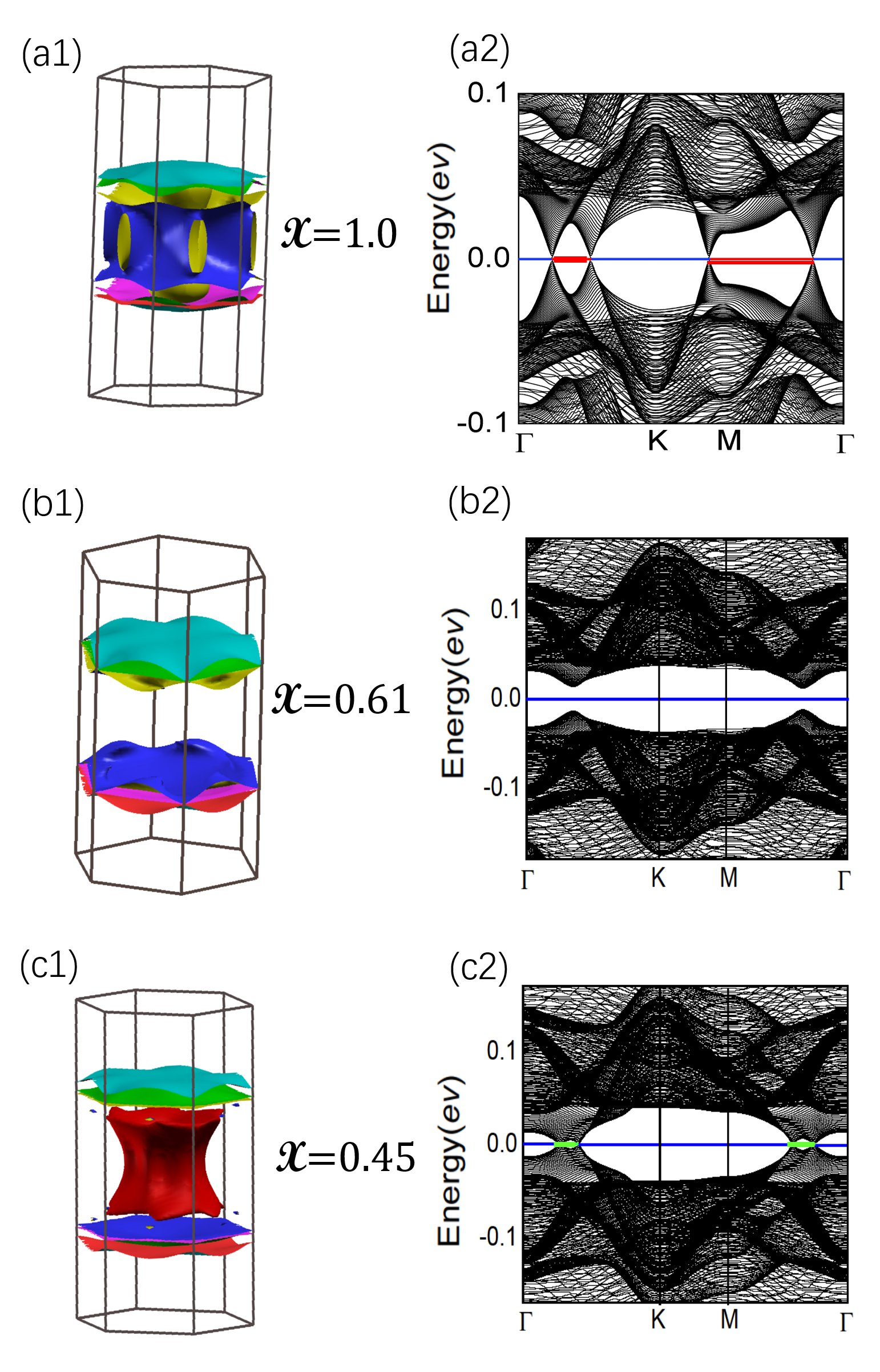}
	\caption
	{(Color online) (a1)-(c1) FSs for KH$_x$Cr$_{3}$As$_{3}$ with  $x$= 1, 0.61 and 0.45. (a2)-(c2) The corresponding surface spectra for the $(001)$ surface. The segment marked red, blue and green represents 4,6 and 8 flat bands respectively.}
	\label{fs2}
\end{figure}

The nontrivial topological invariant $Z(k_x,k_y)$ leads to exactly-flat surface bands on the $(001)$ surface\cite{Liu:033050}. Concretely, for nonzero $Z(k_x,k_y)$, there would be $2Z(k_x,k_y)$ surface bands which are exactly flat when they pass through the momentum $(k_x,k_y)$ in the surface BZ\cite{Liu:033050}. Such knowledge guides us to study the surface spectrum on the $(001)$ surface. For this purpose, we adopt open boundary condition along the $z$- axis and periodic ones along the $x$- and $y$- axes, under which we Fourier transform the BCS-MF Hamiltonian (\ref{BCS}) to the $\left(k_x,k_y,i_z\right)$ space. Then we diagonalize the Bogoliubov-de-Genes equation for the fixed $(k_x,k_y)$ to get the corresponding energy spectrum. The obtained energy spectrum as function of $k_x$ and $k_y$ is shown in Fig.~\ref{edge} along the high symmetric lines in the $(k_{x},k_{y})$ plane. To enhance visibility, we have enlarged the gap amplitude by an order of magnitude. As shown in Fig.~\ref{edge}, some regime in the $(k_x,k_y)$-plane is covered by six flat bands while the remaining regime is covered by four flat bands. We have checked that these flat bands are surface bands as the corresponding wave functions are localized on the $(001)$ surface. Comparing Fig.~\ref{edge} with Fig.~\ref{fs} (b), it's found that the regime within the six Fermi pockets containing the typical $(k_{x2},k_{y2})$ point wherein $Z(k_x,k_y)=2$ is covered by four flat bands, and the remaining regime containing the typical $(k_{x1},k_{y1})$ point wherein $Z(k_x,k_y)=3$ is covered by six flat bands.

The FS topology of  KH$_x$Cr$_3$As$_3$ can be drastically changed upon tuning $x$ which is experimentally accessible\cite{Wu:155108}. On the other hand, Fig.~\ref{pzsoc} shows that the triplet $p_z$-wave pairing in the $S_z=0$ channel is always the leading pairing symmetry, belonging to the SUS-protected TRI topological nodal-line SC with flat surface bands. Consequently, the distribution of the number of flat bands in the surface BZ will be easily engineered through tuning the hydrogen doping. In Fig.~\ref{fs2}, the FSs and surface spectra for $x=1$, $x=0.61$ and $x=0.45$ are shown. Clearly, the different FS topologies for the three different hydrogen-doping levels lead to different distributions of the number of flat bands in the surface BZ. For all the hydrogen-doping levels we studied, the whole surface BZ is covered with flat bands. Particularly in the case of $x=0.61$, due to the simple FS shape with three Q1D FSs, the whole surface BZ is covered with six flat bands.

\section{Discussion and Conclusion}
The SUS-protected TRI topological nodal-line SC with flat surface bands predicted here for the AH${_{\bm x}}$Cr${_{\bm 3}}$As${_{\bm 3}}$ family is similar with that predicted for the A$_{\bm 2}$Cr${_{\bm 3}}$As${_{\bm 3}}$ family, which can be detected in the surface conductance spectrum in the scanning tunneling microscope (STM) experiment by the sharp zero-bias peak\cite{Liu:033050}. There can be inter-site spin-flipping SOC in the two families, which however should be much weaker than the on-site SOC adopted here. In the presence of such weak spin-flipping SOC, the SUS would be slightly broken, leading to slightly broadening of the zero-bias peak in the STM, which is still well detectable\cite{Liu:033050}.

In conclusion, we have performed a RPA-based study on the pairing symmetry and topological properties of the KH$_x$Cr$_3$As$_3$, based on our DFT band structure considering SOC. Our results yield triplet $p_z$-wave pairing in the $(\uparrow\downarrow+\downarrow\uparrow)$ channel to be the leading pairing symmetry. This pairing state belongs to the SUS-protected TRI topological nodal-line SC hosting flat surface bands.  Determined by the $\mathbf{k}$-dependent topological invariant $Z(k_x,k_y)$, the whole $(001)$-surface BZ is covered with topological flat bands, with different regimes covered with different numbers of flat surface bands, as shown by our surface spectra, which can be verified by experiments.

\acknowledgments
We appreciate the help from Li-Da Zhang, Chen Lu and Yu-Bo Liu. This work is supported by the NSFC under the Grant No. 12074031 and 11674025.

\end{document}